\documentclass[reprint, superscriptaddress, amsmath,amssymb, aps, pra, floatfix]{revtex4-1}
\usepackage[T1]{fontenc} 
\usepackage{mathptmx}
\usepackage{graphicx}
\usepackage{dcolumn}
\usepackage{bm}
\usepackage{makecell}
\usepackage[colorlinks, breaklinks, citecolor=blue,linkcolor=black,urlcolor=black]{hyperref}

\begin{document}

\title{Blue-detuned Magneto-optical Trap of BaF molecules}

\author{Zixuan Zeng}
\author{Shoukang Yang}
\author{Shuhua Deng}
\affiliation{Zhejiang Province Key Laboratory of Quantum Technology and Device, School of Physics, and State Key Laboratory for Extreme Photonics and Instrumentation, Zhejiang University, Hangzhou 310027, China
}
\author{Bo Yan}
\email{yanbohang@zju.edu.cn}
\affiliation{%
Zhejiang Province Key Laboratory of Quantum Technology and Device, School of Physics, and State Key Laboratory for Extreme Photonics and Instrumentation, Zhejiang University, Hangzhou 310027, China
}%

\date{\today}

\begin{abstract}
We report the realization of a blue-detuned magneto-optical trap (BDM) of BaF molecules. The $(1+1)$ type BDM and $(1+2)$ type conveyor-belt MOT are explored. While the (1+1) BDM provides only weak trapping force, the conveyor-belt MOT significantly compresses the molecular cloud, achieving a radius of 320 $\mu$m, a temperature of 240 $\mu$K, and a peak density of $1.3\times10^{7}$ cm$^{-3}$, representing a significant improvement over the red MOT. Interestingly, the conveyor-belt MOT of BaF exhibits a large capture velocity, and the loading efficiency from red MOT reaches near unity even without gray molasses. We confirm this by directly loading the laser-slowed molecules into the conveyor-belt MOT.
\end{abstract}

\maketitle


Achieving a quantum gas of polar molecules is a significant goal in cold molecule physics \cite{Langen2024, Zeng2025}. While degenerate Fermi gases and Bose-Einstein condensates have been realized using ultracold atom-associated molecules like KRb \cite{DeMarco2019}, NaK \cite{Schindewolf2022}, and NaCs \cite{Bigagli2024}, creating such a quantum gas with directly laser-cooled molecules remains an open challenge. Many well-established cooling techniques from cold atom physics, such as Raman sideband cooling \cite{Caldwell2020, Lu2024,  Bao2024} and evaporative cooling \cite{Schindewolf2022, Bigagli2024}, offer potential avenues for addressing this challenge. A primary obstacle is the low phase space density (PSD) of molecular magneto-optical traps (MOTs) \cite{Barry2014, Anderegg2017, Collopy2018, Vilas2022, Zeng2024, Lasner2024, PadillaCastillo2025}, typically around $10^{-13}$, which is much smaller than the atomic case ($\sim 10^{-6}$). Improving the MOT's PSD is crucial for efficient loading molecules into conservative traps, such as optical dipole traps \cite{Anderegg2018} or optical tweezers \cite{Anderegg2019}, which is essential for achieving quantum molecular gas, and enabling studies of molecules in quantum information \cite{Bao2023, Holland2023} and quantum chemistry \cite{Cheuk2020, Anderegg2021}.

The use of Type-II transitions in laser cooling of molecules, which is necessary for establishing a cycling transition condition for rotational states \cite{Stuhl2008}, presents a challenge to achieve high PSD in a molecular MOT. This transition type tends to cause heating near zero velocity \cite{Tarbutt2015}, resulting in weak and hot MOTs with typical densities of $10^5$ cm$^{-3}$ and temperatures in the millikelvin range. Gray molasses techniques, which utilize blue-detuned light and position-dependent dark states to implement  Sisyphus cooling, offer a way to circumvent this limitation \cite{Weidemueller1994, Fernandes2012}. These techniques can cool molecules to the $50\mu$K level \cite{Truppe2017b, Anderegg2018, McCarron2018}, and further cooled to a few $\mu$K with $\Lambda-$enhanced gray molasses\cite{Cheuk2018, Caldwell2019, Ding2020, Hallas2023}. However, gray molasses do not provide trapping capabilities. To achieve compression in both spatial and velocity space, the blue-detuned MOT (BDM) was developed \cite{Jarvis2018}, and has been demonstrated with molecules such as YO \cite{Burau2023}, CaF\cite{Li2024}, SrF\cite{Jorapur2024}, increasing their PSDs by two to three orders of magnitude. Furthermore, a conveyor-belt MOT, a novel BDM using moving lattices to enhance trapping force\cite{Li2025}, has been realized with CaOH \cite{Hallas2024} and CaF \cite{Yu2024} molecules, achieving a PSD of $2.4\times 10^{-6}$ for CaF.

\begin{figure}[t]
\centering
\includegraphics[width=0.46\textwidth]{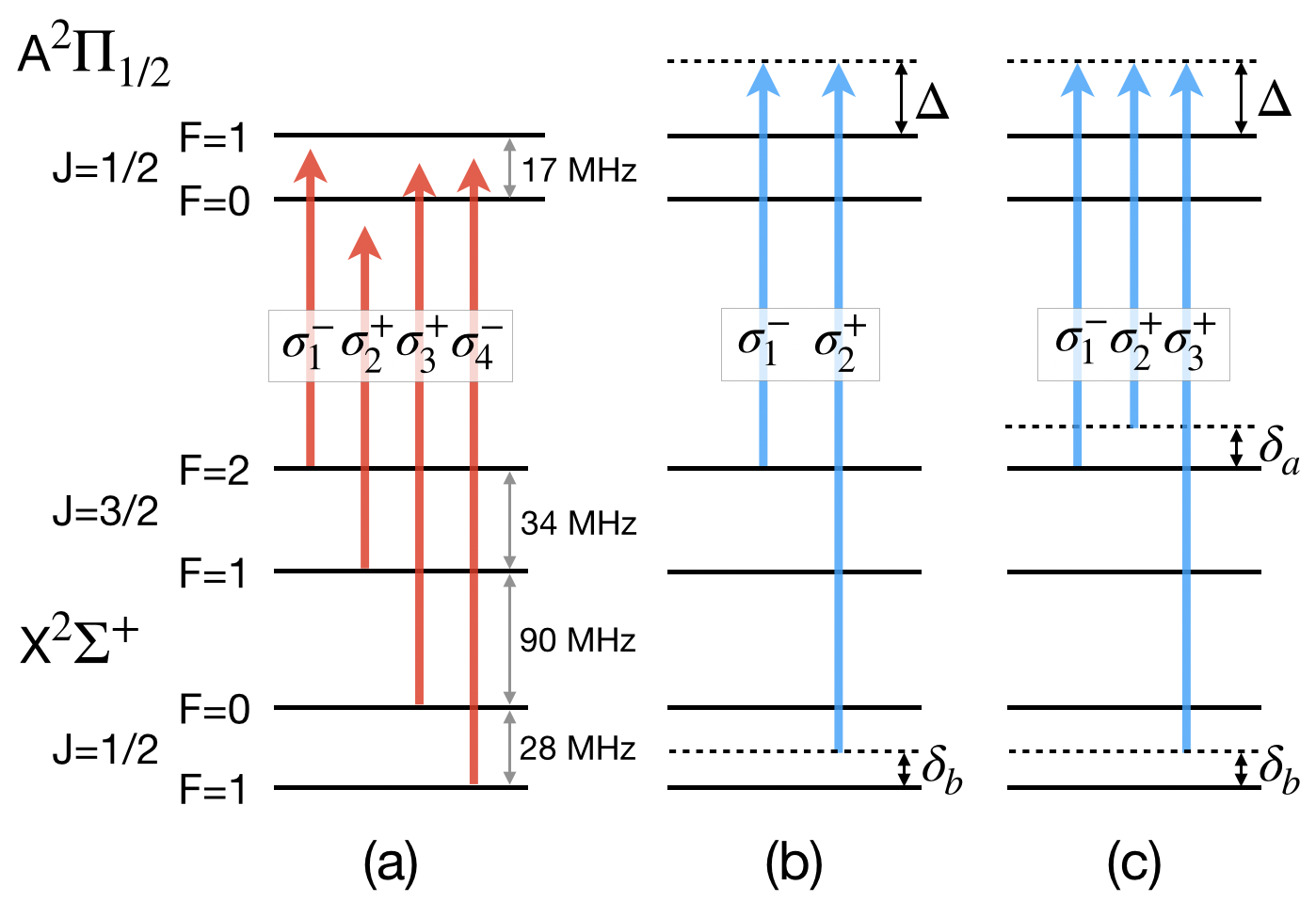}
\caption{(color online) \label{level}
The energy levels involved with BaF molecules for (a) red MOT, (b) (1+1) BDM, and (c) (1+2) conveyor-belt MOT. We define the frequency difference between the $\sigma_1$ light and the $|J=3/2, F=2\rangle$ to $|J'=1/2, F'=1\rangle$ transition as $\Delta$ for all three cases. The frequency differences between $\sigma_1$ and the other lights are specified as follows: (a) red MOT $\{0,6,122,148\}$ MHz (corresponding to $\sigma_1$, $\sigma_2$, $\sigma_3$, $\sigma_4$), (b) (1+1) BDM $\{0,152-\delta_b\}$ MHz ($\sigma_1$ and $\sigma_2$), (c) (1+2) conveyor-belt MOT $\{0,-\delta_a,152-\delta_b\}$ MHz ($\sigma_1$, $\sigma_2$, $\sigma_3$). The polarization of each light is indicated in the plot.
}
\end{figure}

Here, we report the realization and detailed study of a BDM for BaF molecules. BaF, the heaviest molecule trapped in a MOT to date \cite{Zeng2024}, holds significant promise for precision measurements, such as searches for the electron electric dipole moment (eEDM) \cite{Chen_2016, Aggarwal2018}. Compared to other laser-cooled molecules, the heavier mass, longer wavelength transition, and narrower excited state linewidth of BaF result in weaker laser cooling and trapping. In addition, BaF possesses a large g-factor and hyperfine splitting (17 MHz $\sim6\Gamma$) in its excited states compared to other molecules   \cite{Bu2022}. These unique characteristics introduce new technical challenges and necessitate a refined search regime for achieving a good BDM.

The experimental setup is similar to that described in our previous works \cite{Bu2017, Zhang2022, Zeng2024}. BaF molecules are created via laser ablation and cryogenically cooled by a 4K Helium buffer gas. After exiting the buffer gas cell, the BaF molecular beam is slowed by a frequency-chirped slowing laser, resulting in a population of molecules with near-zero velocity suitable for trapping in a MOT.  The molecules are then captured by a red MOT, typically yielding $1\times 10^4$ molecules at a temperature of approximately 3 mK. The red MOT has a size of 1.4(1) mm and a peak density of $2.5\times10^{5}$ cm$^{-3}$. The laser configuration for the main cooling transition of the red MOT is illustrated in Fig. 1(a). The peak intensity of each MOT beam is $\sim$11 mW/cm$^2$ and the axial magnetic field gradient is $ B_z^\prime=$8 G/cm.

\begin{figure}[t]
\centering
\includegraphics[width=0.45\textwidth]{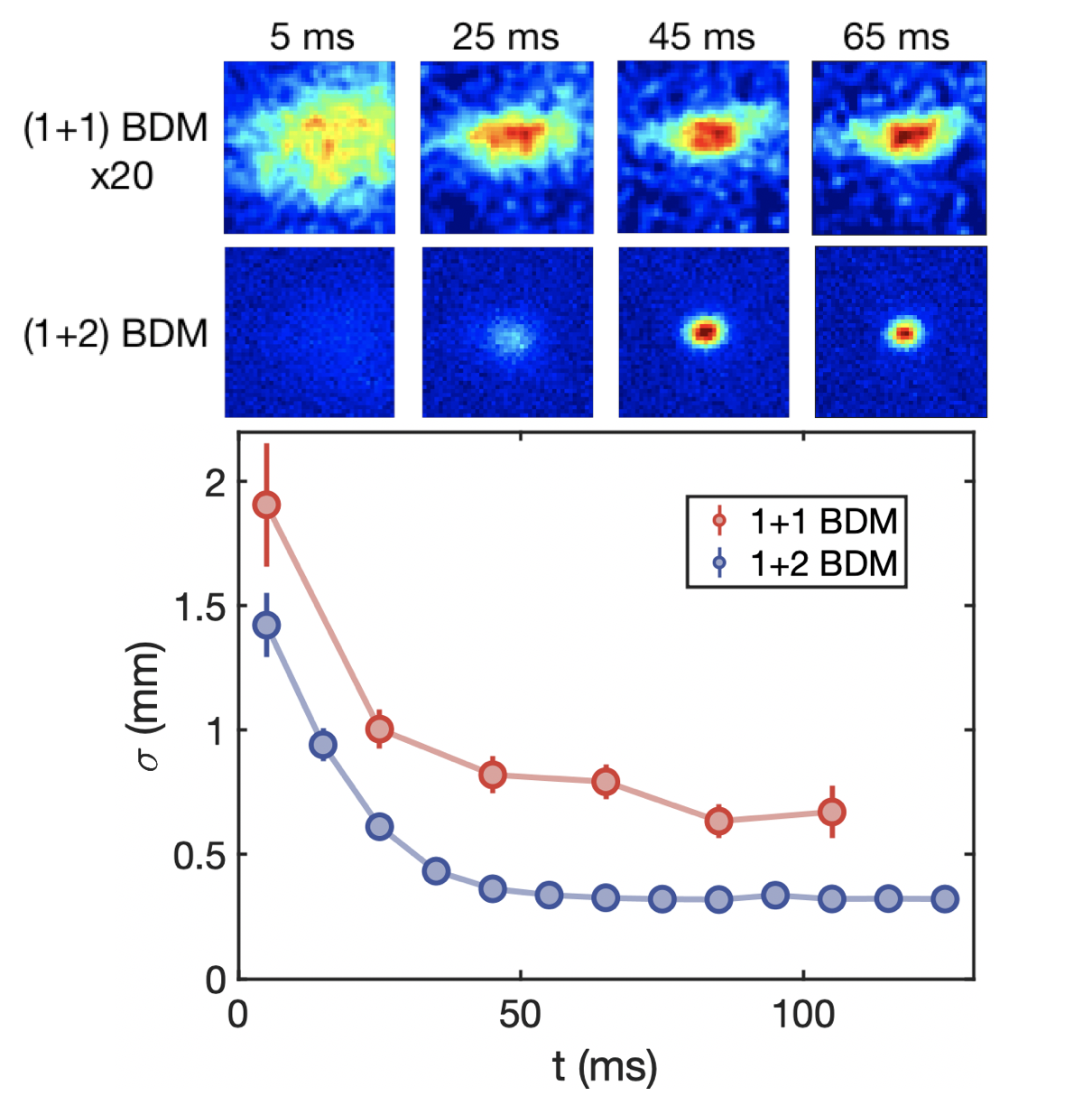}
\caption{(color online) \label{compress}
The compression process of both the (1+1) BDM and the (1+2) conveyor-belt MOT. The top panel displays the in-situ images of the molecular cloud with a 10 ms exposure at various time points; note that different color scales are used for these two cases: the signal of (1+1) BDM is amplified 20 times for better visibility. The bottom panel illustrates the cloud compression, quantified by the geometry-averaged Gaussian radius $\sigma = \sigma_{axial}^{1/3}\times\sigma_{radial}^{2/3}$ obtained from a 2D Gaussian fitting.
}
\end{figure}

Following a 20~ms loading period into the red MOT, the molecules are subjected to BDM compression. We investigated two BDM laser configurations, shown schematically in Fig.1(b) and (c). The $(1+1)$ BDM configuration, depicted in Fig.1(b), has been successfully implemented in previous experiments with YO and CaF molecules \cite{Burau2023, Li2024}. Optimizing this configuration in our experiment resulted in $\Delta=$ 10 MHz, $\delta_b=$ 2.7 MHz. Two lasers have polarizations of $\{\sigma_1-, \sigma_2+\}$ (Fig.1(b)), and the peak intensity of both frequency components is 10 mW$/$cm$^2$. With these, the molecular cloud was compressed to 630(70) $\mu m$ and cooled to 190(100) $\mu K$. This represents an improvement over the red-MOT alone, but the density remained low. A similar limitation has been reported for SrF molecules, where the $(1+1)$ BDM configuration yielded sub-optimal results\cite{Jorapur2024}. 

We attribute this suboptimal performance to BaF's relatively high mass and the $g$-factor of ground states. Effective BDM compression of molecular ensembles requires both sub-Doppler cooling and a significant trapping force. CaF molecules, due to their lighter mass, can achieve a sufficient trapping force without a dual-frequency mechanism \cite{Li2024}. YO molecules, with their near-zero $g$-factor in their magnetic sublevels, maintain effective sub-Doppler cooling even in a gradient magnetic field \cite{Burau2023}. Therefore, although it requires a compression time on the order of 100 ms, a small size can ultimately be achieved. In contrast, BaF does not possess the aforementioned features, which we believe is the reason why we only observed weak trapping and failed to achieve optimal compression in the $(1+1)$ BDM configuration.

\begin{figure}[t]
\centering
\includegraphics[width=0.48\textwidth]{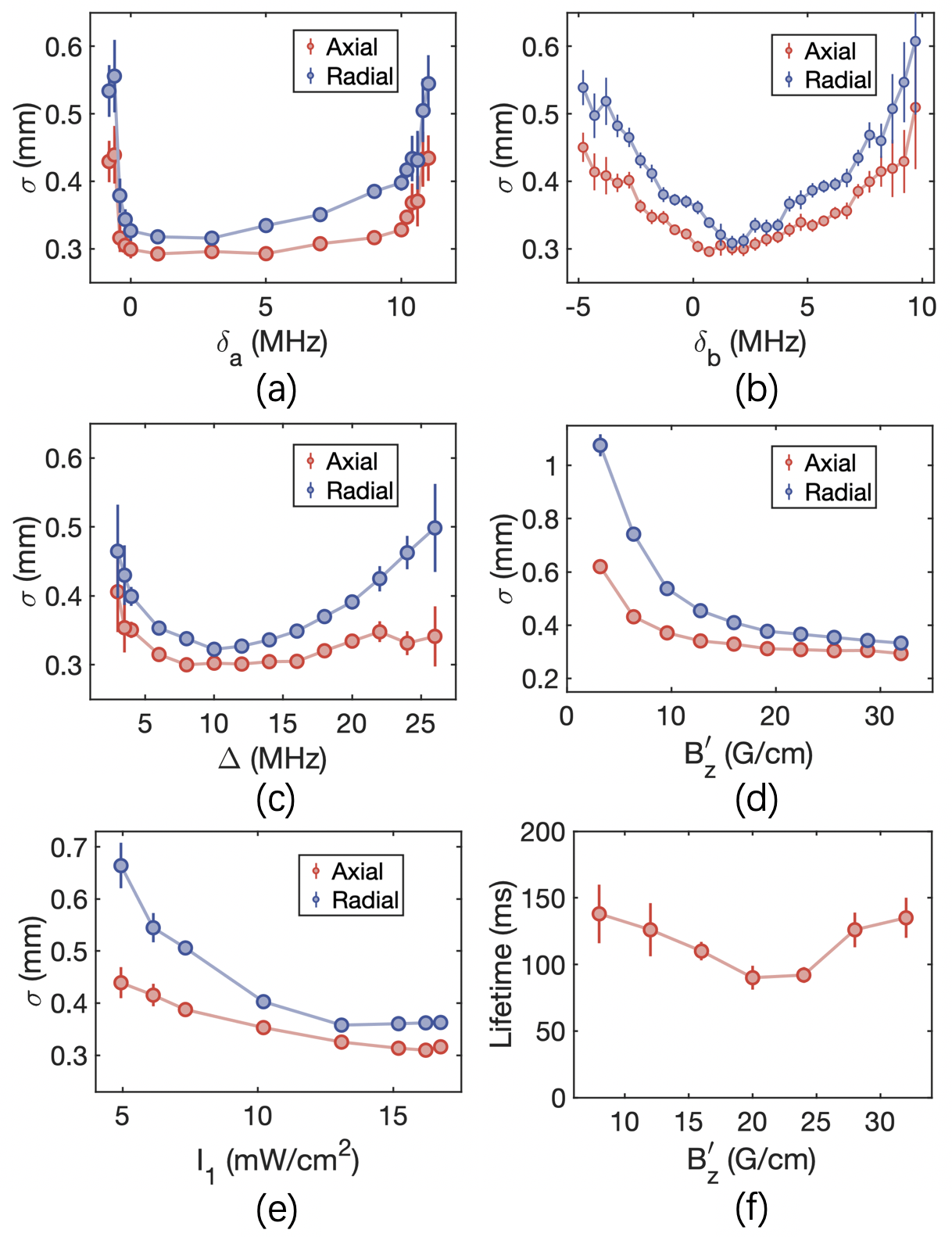}
\caption{(color online) \label{blue_MOT}
The size of the conveyor-belt MOT depends on: (a) two-photon offset detuning $\delta_a$, (b) two-photon detuning $\delta_b$, (c) single-photon detuning $\Delta$, (d) $z$-direction magnetic field gradient $B_z^{\prime}$, and (e) beam intensity $I_1$. Additionally, the lifetime dependence on $B_z^{\prime}$ is shown in (f). The radial and axial sizes are represented by red and blue circles, respectively. 
}
\end{figure}

To address this limitation, we implemented a $(1+2)$ type BDM configuration, which was identified as a moving lattice technique, termed "conveyor-belt MOT" \cite{Li2025, Yu2024}. The laser configuration for the conveyor-belt MOT is shown in Fig. 1(c). This configuration utilizes two frequency-offset lasers $\sigma_1^-$ and $\sigma_2^+$, separated by a small frequency difference $\delta_a$. The superposition of these lasers with their retro-reflected beams creates two moving optical lattices with opposite velocities $\pm \delta_a/k$. The type-II transition in molecular cooling preferentially pumps molecules to the moving lattice towards the center of the MOT, resulting in a significant increase in the trapping force.

Figure 2 illustrates the compression process transitioning from a red MOT to a BDM. When we initiate the BDM compression, the main laser configuration is switched from the red MOT to the BDM configuration instantaneously  (set as $t=0$ ms) and then kept unchanged.  The top panels show images of the molecular cloud at various time points for both (1+1) and (1+2) type BDM configurations. For the (1+1) case, the detunings are $\{\Delta,~\delta_b\}=\{10,~ 2.7\}$ MHz, $I_{\sigma_1^-}=I_{\sigma_2^+}=10$~mW$/$cm$^2$ and the magnetic filed gradient stays at 8 G/cm. Further increase of the gradient leads to a significant decrease in the number of molecules. For the (1+2) case,  $\{\Delta, ~\delta_a, ~\delta_b\}=\{10,~3,~2.2\}$ MHz, $\{I_{\sigma_1^-}, ~I_{\sigma_2^+},~ I_{\sigma_3^+}\}=\{13, ~8, ~13\}$~mW$/$cm$^2$ and the magnetic field gradient is ramped from 8 G/cm to 32 G/cm in 40 ms. These images were acquired using an sCMOS camera with a 10 ms in-situ exposure. The bottom plot shows the fitted MOT sizes as a function of BDM compression time.  Clearly, the conveyor-belt MOT provides significantly improved compression. Both the final size and compression time are much better for the (1+2) case. Finally, the optimized conveyor-belt MOT achieved a temperature of  240(60) $\mu$K and a cloud size of 320(20) $\mu$m. The peak density achieved was $1.3\times10^7$ cm$^{-3}$, which is a 50 times improvement compared with the red MOT. In our experiment, compressing the red MOT to the conveyor-belt MOT takes approximately 50 ms. This timescale is longer than that observed for SrF \cite{Jorapur2024} and CaF \cite{Yu2024}. This difference likely arises from the lower recoil velocity of BaF (3 mm/s) compared to CaF and SrF, leading to a slower compression process and a larger ultimate MOT size.

Figure 3 presents the systematic characterization of our conveyor-belt MOT. 
In Fig.3(a), we explore the effect of varying the two-photon offset detuning of $\delta_a$, which influences the velocity of the moving lattice, when $\{\Delta, ~\delta_b\}= \{10,~2.2\}$ MHz and  $B_z'$=32 G/cm. The conveyor-belt MOT demonstrates robust performance across a relatively broad range of detunings, $\delta_a=0-10$~MHz. This range is constrained by the hyperfine structures of BaF molecules. When $\delta_a=0$~MHz, the effects of $\sigma_1-$ and $\sigma_2+$ cancel each other out and offer zero trapping force. When $\delta_a=\Delta=10$~MHz, $\sigma_2+$ is on resonance with the $|J=3/2,F=2\rangle\rightarrow|J'=1/2,F'=1\rangle$ transition and results in a significant heating.

Figure 3(b) illustrates the relationship between MOT size and the two-photon detuning of $\delta_b$, when $\{\Delta, ~\delta_a\}= \{10,~3\}$ MHz and  $B_z'$=32 G/cm. Optimized performance is achieved at $\delta_b=2.2$ MHz. Similar to $\delta_a$, conveyor-belt MOT can work within a $\delta_b$ range of about 10 MHz, even when $\delta_b$ is less than zero, which shows a clear distinction from traditional (1+1) BDM.

Figure 3(c) illustrates the dependence of MOT size on the single-photon detuning of $\Delta$, when  $\{\delta_a, ~\delta_b\}= \{3,~2.2\}$ MHz and  $B_z'$=32 G/cm. The optimized compression is observed around $\Delta\simeq 10$~MHz, with similar behavior also apparent around 10 MHz.  These results are consistent with previous findings using (1+1) type BDM. When $\Delta$ approaches 3 MHz, thus becoming comparable to $\delta_a$ and $\delta_b$, the sub-Doppler cooling from $\sigma_2+$ and $\sigma_3+$ lasers transitions to heating, leading to a rapid increase in MOT size. Conversely, while increasing $\Delta$ enhances sub-Doppler cooling, the reduced molecular scattering rate from the larger detuning weakens the trapping force and increases the MOT size.

\begin{figure}[t]
\centering
\includegraphics[width=0.45\textwidth]{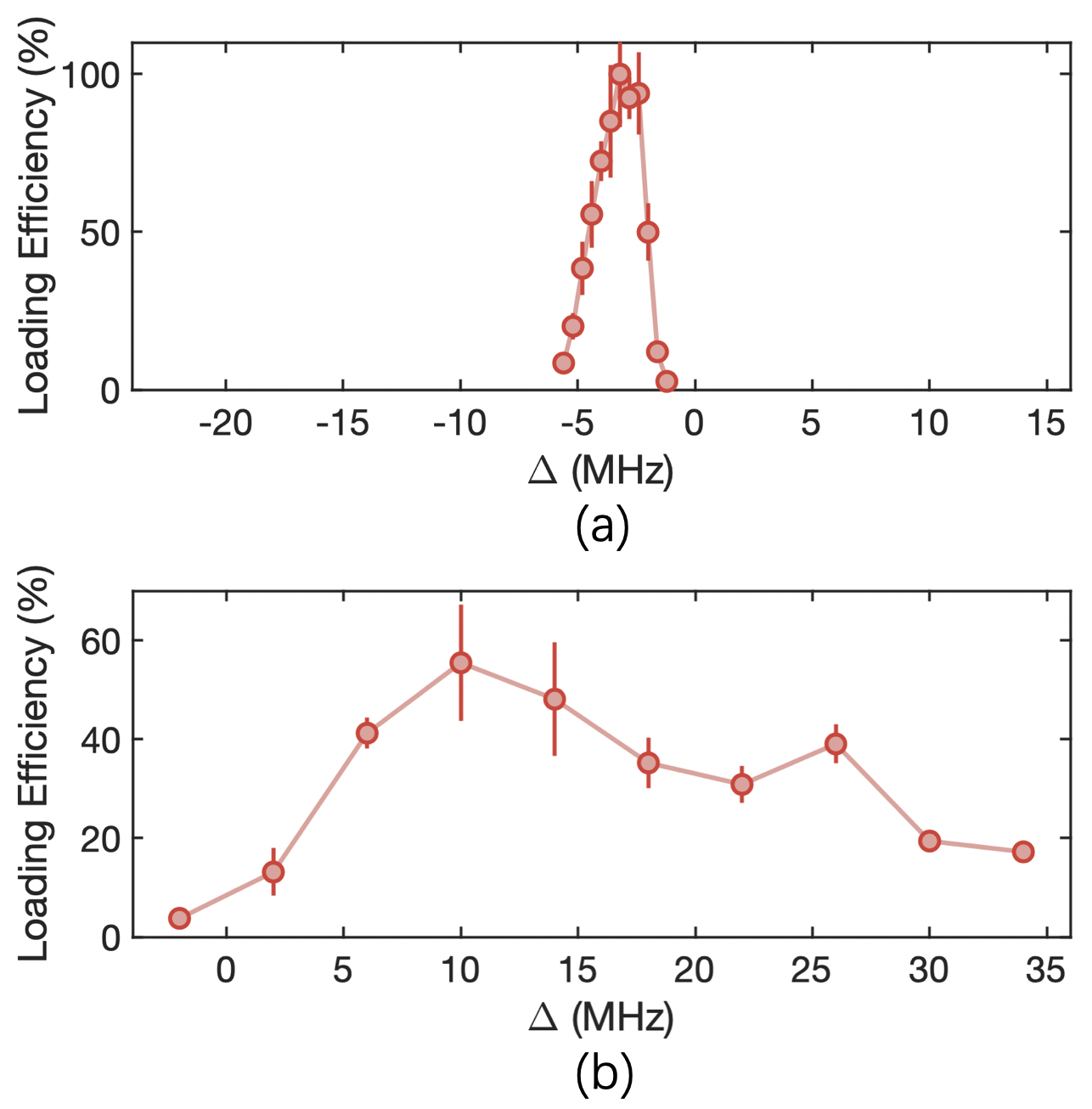}
\caption{(color online) \label{Dloading}
The loading efficiency as a function of the single-photon detuning $\Delta$  when directly loading laser-slowed BaF molecules into (a) a red MOT and (b) a conveyor-belt MOT. A much larger detuning window shows up for case (b).
}
\end{figure}

Figure 3(d) demonstrates the compression of the conveyor-belt MOT by increasing the magnetic field gradient. 
Efficient compression occurs below 10 G/cm. 
Although the MOT size continues to decrease up to the maximum gradient of 32 G/cm (limited by our coil's current), the compression rate slows significantly beyond 20 G/cm. 

Figure 3(e) demonstrates the dependence of MOT size on the laser beam intensity. We set the beam intensities of $\sigma_1-$ and $\sigma_3+$ as equal to each other and denote them as $I_1$, keep the beam intensity of $\sigma_2+$ light at 8 mW/cm$^2$ constant, and measure the variation of MOT size under different $I_1$. 
Contrary to the expected case where $I_1$ equals $I_{\sigma_2+}$, the minimum size is achieved when $I_1$ is approximately twice $I_{\sigma_2+}$, which indicates that although the conveyor-belt effect is crucial for MOT trapping, the compression of the molecular cloud results from the combined action of multiple mechanisms. As observed in (1+1) BDM, the combination of $\sigma_1-$ and $\sigma_3+$ can also provide part of the cooling and trapping force.

Figure 3(f) demonstrates the dependence of the MOT lifetime on the magnetic field gradient. Unlike the MOT lifetime of CaOH reported in \cite{Hallas2024}, the lifetime of BaF conveyor-belt MOT does not appreciably decrease with the increase of magnetic field gradient. It can maintain a lifetime of more than 100 ms in most parameter ranges, which is about the same as the red MOT lifetime under a similar scattering rate. The conveyor-belt MOT scattering rate is measured at 32 G/cm by turning off the 898 nm $(v=2, N=1)$ repump light, and the measurement yields a value of $5(1)\times10^5$ s$^{-1}$.

One intriguing observation is that the capture velocity of the conveyor-belt MOT for BaF molecules is comparable to that of the red MOT. This suggests the possibility and is verified in our experiment of directly loading laser-slowed BaF molecules into the conveyor-belt MOT, bypassing the red MOT. To quantify loading efficiency, we illuminate both the directly loaded BDM and red MOT with the same red MOT beam, defining the ratio of their fluorescence signals as the loading efficiency. As shown in Fig.4, the red MOT exhibits a narrow loading detuning window (approximately 3~MHz), whereas the conveyor-belt MOT, although achieving about half the molecular number, possesses a significantly broader window (approximately 20 MHz). This finding is consistent with our results in Fig.2 (c). Given the larger detuning window and better signal-to-noise ratio, the BaF conveyor-belt MOT could be potentially used as an improved tool for initial MOT searching.

To conclude,  we have successfully realized both the (1+1) and (1+2) type BDM for BaF molecules. Due to BaF's heavy mass, relatively long wavelength transition, and narrow excited state, the (1+1) type BDM yields a weak trapping force and a poor compression of the molecule cloud. Fortunately, the (1+2) type conveyor-belt MOT provides a much stronger trapping and cooling force. This method effectively compresses the molecular cloud by a factor of 4.5 and reduces the molecular temperature by more than one order of magnitude. Although this result is not yet on par with lighter molecules such as YO, CaF, and SrF \cite{Burau2023, Li2024, Yu2024, Jorapur2024}, it significantly improves trapped heavy molecules. 

With this advancement, it is now feasible to attempt loading BaF to a conservative trap, such as the dipole trap or optical lattice. Implementing $\Lambda-$enhanced gray molasses within these traps should further improve BaF's PSD, enabling valuable applications in quantum information and precision measurement with cold molecules. 


We acknowledge the support from the National Natural Science Foundation of China under Grant No. 12425408 and No. U21A20437, the National Key Research and Development Program of China under Grant No.2023YFA1406703 and No. 2022YFA1404203.


\bibliographystyle{apsrev4-1}
\bibliography{blue_MOT0522}

\end{document}